\documentclass[11pt]{article}
\usepackage{amsmath,amssymb,color,graphics,epsfig}

\usepackage{amsfonts}

\newcommand{\be}{\begin{equation}}
\newcommand{\ee}{\end{equation}}
\newcommand{\bea}{\setlength\arraycolsep{2pt} \begin{eqnarray}}
\newcommand{\eea}{\end{eqnarray}}

\def\ft#1#2{{\textstyle{\frac{\scriptstyle #1}{\scriptstyle #2} } }}
\def\fft#1#2{{\frac{#1}{#2}}}

\def\0{{\sst{(0)}}}
\def\1{{\sst{(1)}}}
\def\2{{\sst{(2)}}}
\def\3{{\sst{(3)}}}
\def\4{{\sst{(4)}}}
\def\5{{\sst{(5)}}}
\def\6{{\sst{(6)}}}
\def\7{{\sst{(7)}}}
\def\8{{\sst{(8)}}}
\def\sst#1{{\scriptscriptstyle #1}}

\begin{document}

\begin{flushright}
\end{flushright}

\vspace{25pt}
\begin{center}
{\large {\bf Charged Dilatonic AdS Black Holes and Magnetic AdS$_{D-2}\times R^2$ Vacua}}

\vspace{10pt}
H. L\"u

\vspace{10pt}

{\it Department of Physics, Beijing Normal University,
Beijing 100875, China}

\vspace{40pt}

\underline{ABSTRACT}
\end{center}

We consider $D$-dimensional Einstein gravity coupled to two $U(1)$ fields and a dilaton with a scalar potential.  We derive the condition that the analytical AdS black holes with two independent charges can be constructed. Turning off the cosmological constant, the extremal Reissner-Nordstr\o m black hole emerges as the harmonic superposition of the two $U(1)$ building blocks. With the non-vanishing cosmological constant, our extremal solutions contain the near-horizon geometry of AdS$_2\times R^{D-2}$  with or without a hyperscaling. We also obtain the magnetic AdS$_{D-2}\times {\cal Y}^2$ vacua where ${\cal Y}^2$ can be $R^2$, $S^2$ or hyperbolic 2-space.  These vacua arise as the fix points of some super potentials and recover the known supersymmetric vacua when the theory can be embedded in gauged supergravities.  The AdS$_{D-2}\times R^2$ vacua are of particular interest since they are dual to some quantum field theories at the lowest Landau level. By studying the embedding of some of these solutions in the string and M-theory, we find that the M2/M5-system with the equal M2 and M5 charges can intersect with another such M2/M5 on to a dyonic black hole.  Analogous intersection rule applies also to the D1/D5-system.  The intersections are non-supersymmetric but in the manner of harmonic superpositions.

\vfill {\footnotesize Emails: mrhonglu@gmail.com}

\thispagestyle{empty}

\pagebreak



\newpage

\section{Introduction}

Analytic solutions of electrically-charged black holes that are asymptotic to anti-de Sitter (AdS) space-times are useful gravitational backgrounds to study some strongly coupled dual field theories on the AdS boundaries.  The most celebrated examples are the Reissner-Nordstr\o m (RN) AdS black holes of the cosmological Einstein-Maxwell theories in general dimensions.  However, only in four and five dimensions, Einstein-Maxwell gravities can be embedded in supergravities and hence in the string or M-theory. The phenomenological AdS/CFT correspondence of Einstein-Maxwell gravities in higher than five dimensions is based on holographic principles that are beyond any known consistent fundamental theory.

Charged dilatonic AdS black holes are much more difficult to construct.  Most of the known examples are those of supergravities.  The examples include the $U(1)^3$-charged and $U(1)^4$-charged AdS black holes of gauged supergravities in five \cite{Behrndt:1998jd} and four \cite{Duff:1999gh} dimensions, and also charged solutions in seven \cite{tenauthor} and six \cite{Cvetic:1999un} dimensions.  (See also \cite{Klemm:2012yg,Klemm:2012vm}.  Many examples of charged rotating AdS black holes in gauged supergravities have been constructed, including the general rotating charged black holes in five-dimensional minimal gauged supergravity \cite{Chong:2005hr} and  $U(1)^3$ gauged supergravity \cite{Wu:2011gq}. Other notable examples include charged rotating black holes in ${\cal N}=4$, $D=4$ $SO(4)$ \cite{Chong:2004na}, ${\cal N}=1$, $D=7$ \cite{Chow:2007ts}, and ${\cal N}=(1,1)$, $D=6$ \cite{Chow:2008ip} gauged supergravities.)
In four and five dimensions, when all the charges are set to equal, the solutions become the RN-AdS black holes.  In ungauged supergravities, the extremal RN black holes can be viewed as bound states of the basic $U(1)$ building blocks with zero binding energy \cite{Rahmfeld:1995fm}.  They can be lifted and become harmonic superpositions of intersecting M-branes or D$p$-branes. (See, {\it e.g.}, \cite{Tseytlin:1996bh}.)

    The goal of this paper is to generalize the above supergravity
phenomenon and find theories with these properties in general dimensions.  In constructing solutions in supergravities, higher-order fermionic terms play absolute no role.  The useful property is the existence of consistent Killing-spinor equations.  By consistent, we mean that the integrability conditions under some specific $\Gamma$-matrix projections are automatically satisfied by the full set of bosonic equations of motion.  This property exists for all on-shell supergravities.  It was shown in \cite{Lu:2011nga,Lu:2011zx,Lu:2011vk,Liu:2012jra} that such consistent Killing-spinor equations can exist also for some bosonic theories that cannot be supersymmetrized, nor embeddable in any supergravity.
There are then pseudo-supergravity extensions of these theories whose full actions are invariant under the pseudo-supersymmetric transformation rules up to and including the quadratic order in fermions, but not higher orders. (See, for examples, \cite{Lu:2011ku,Liu:2011ve}.)  It was shown in \cite{Liu:2011ve} that the Kaluza-Klein theory in any dimensions can be pseudo-supersymmetrized.  The gauging of the Kaluza-Klein vector induces a scalar potential with the AdS fixed point. The charged Kaluza-Klein AdS black hole in {\it any} dimensions can be obtained \cite{Liu:2012jra}, which is the special case of the two-charge AdS black holes \cite{Chow:2011fh}.
These are the only known examples of charged dilatonic AdS black holes in general dimensions.  Furthermore, the general charged rotating AdS black holes for this theory was also obtained \cite{Wu:2011zzh}.  In these pseudo-supergravities, which exist in general dimensions, the construction of the extremal $p$-branes is identical to that of the BPS $p$-branes in supergravities.  The $p$-branes can intersect in the manner of harmonic superpositions just as in usual supergravities.

The simplest theory in which the RN black hole may emerge as a bound state involves two basic ingredients. Thus our theory consists of the metric, a dilaton $\phi$ and two Maxwell fields $(A_1,A_2)$, with the Lagrangian
\begin{equation}
e^{-1} {\cal L}_D = R - \ft12 (\partial\phi)^2 - \ft14 e^{a_1\phi} F_1^2 - \ft14 e^{a_2\phi} F_2^2 - V(\phi)\,,\label{lag1}
\end{equation}
where $e=\sqrt{-g}$ and $F_i=dA_i$ and $(a_1, a_2)$ are dilaton coupling constants. It is clear that if the scalar potential has a fixed point and $a_1 a_2<0$, the dilaton can be decoupled and the theory reduces to the Einstein-Maxwell theory with a cosmological constant. Static solutions involving one vector field and some scalar potentials with unusual asymptotics were obtained in \cite{Chan:1995fr}.  For our purpose a black hole is defined to have both the horizon and a maximally-symmetric asymptotic space-time.

     In section 2, we study the theory with no scalar potential.  It was
shown in \cite{stainless} that one can always construct a $p$-brane when only a single field strength is turned on.  The construction involving multiple field strengths is much subtler and requires some specific choice of dilaton couplings.  We find that if the constants $(a_1,a_2)$ satisfy
\begin{equation}
a_1 a_2 = - \fft{2 (D-3)}{D-2}\,,\label{a1a2cons}
\end{equation}
analytical black holes with two independent charges can be obtained.  (See the book \cite{Ortin:2004ms} on the construction of such solutions.)
Furthermore, extremal RN black holes emerge in our theories as bound states indeed of the two basic extremal black holes associated with $A_i$, with zero binding energy.  The general solutions contain the supersymmetric examples in four and five dimensions.

In section 3, we consider adding a scalar potential, while keeping the condition (\ref{a1a2cons}). We derive the scalar potential such that analytical charged AdS black holes can be obtained.  The scalar turns out to be always massless (in the AdS sense) and the scalar potential can be expressed in terms of a super potential.  The charged AdS black holes have the similar structure as those in gauged supergravities.  Depending on the charge configurations, taking the extremal limit gives rise to the AdS$_2\times R^{D-2}$ near-horizon geometry with or without a hyperscaling.

In section 4, we consider the special case with $a_1=-a_2$ in four dimensions.  The two-charge solutions can be easily mapped to the dyonic black hole and the two electric charges of different field strengths become the electric and magnetic charges of the same field strength.  We find that the solution can be embedded in the string and M-theory.  Lifting the solutions back to higher dimensions, we find that they describe intersections of an M2/M5-brane of equal M2 and M5 charges
with another such M2/M5 system.  The intersection is in the manner of harmonic superpositions, but non-supersymmetric.  The same story applies to the D1/D5-system.  This provides a new embedding of the four-dimensional RN black hole in the string and M-theory.

In section 5, we consider the AdS$_{D-2}\times {\cal Y}^2$ vacua that carry magnetic 2-form fluxes. These solutions can arise as the near-horizon geometry of some extremal magnetic $(D-4)$-branes. The two-dimensional internal space ${\cal Y}^2$ can be $R^2$, $S^2$ or hyperbolic space.  We use the super potential method and show that the vacua may emerge as fixed points of the super potentials.  When the theory can be embedded in gauged supergravities, we recover the known supersymmetric vacua.  The magnetic AdS$_{D-2}\times R^2$ vacua are of particular interest since they are expected to be dual to some quantum field theories at the lowest Landau level \cite{Almuhairi:2011ws}.

We conclude the paper in section 6.

\section{Charged asymptotic-flat black holes}

In this section, we set the scalar potential $V(\phi)$ to zero in the Lagrangian (\ref{lag1}), and concentrate on charged black holes that are asymptotic flat. If we turn on both $A_i$ independently, the theory for general $(a_1,a_2)$ does not admit analytic black holes.  We shall determine the condition on $(a_1,a_2)$ so that the system will give such analytical solutions.  It is advantageous for later purpose that we reparameterize these dilaton coupling constants as
\begin{equation}
a_1^2=\ft{4}{N_1} - \ft{2 (D-3)}{D-2}\,,\qquad a_2^2 = \ft{4}{N_2}- \ft{2(D-3)}{D-2}\,.\label{Nidef}
\end{equation}
The reality condition of each $a_i$ requires that
\begin{equation}
0<N_i \le \ft{2(D-2)}{D-3}\,.
\end{equation}
Note that $N_i=1,2$ satisfy the above condition in general dimensions. If both $N_i$'s are outside the range, the Lagrangian can still be made real by letting $\phi\rightarrow {\rm i}\phi$, corresponding to having a ghost-like dilaton.  We shall not consider such situation at all.

In supergravities, the quantities $N_i$ are typically positive integers. A black hole or a $p$-brane with such $N_i=1$ can be interpreted as the most basic building block.  Examples include the M-branes, all the D-branes, and the Kaluza-Klein black holes. The $N=2$ solution may emerge as the bound state of such two $N=1$ states.  As we shall see later, the zero binding energy implies that the resulting $N$ is given by $N_1 + N_2$. In this paper, we shall relax the condition that $N_i$ can only be integers and generalize the above phenomenon of $p$-brane superpositions which is common in supergravities to the theory (\ref{lag1}) in general dimensions.

The spherically-symmetric and static ansatz is given by
\begin{equation}
ds^2=-e^{2A} dt^2 + e^{2B} (dr^2 + r^2 d\Omega_{D-2}^2)\,,\qquad
A_i = e^{C_i} dt\,.
\end{equation}
The functions $(A,B,C_i)$ and the dilaton $\phi$ all depend on the radial coordinate $r$ only.  The equations of motion for the vector fields imply that
\begin{equation}
(e^{C_i})' = \fft{\lambda_i}{r^{D-2}} e^{A - (D-3) B - a_i \phi}\,,
\end{equation}
where a prime denotes a derivative with respect to $r$. Using the useful formulae obtained in \cite{stainless,toda}, one finds that the function $X\equiv A + (D-3)B$ can be solved straightforwardly, given by \cite{toda}
\begin{equation}
e^X=1 - k \rho^2\,,
\end{equation}
where $k$ is the integration constant and $\rho=r^{3-D}$.  Defining a new coordinate $\xi$ such that $d\xi=e^{-X}d\rho$, the equations can be reduced to
\begin{eqnarray}
\ddot \phi &=& -\ft1{2(D-3)^2} (\lambda_1^2 a_1 e^{-a_1 \phi} +\lambda_2^2 a_2 e^{-a_2\phi})e^{2A}\,,\cr
\ddot A &=& \ft{1}{2(D-2)(D-3)} ( \lambda_1^2 e^{-a_1\phi} + \lambda_2^2 e^{-a_2\phi}) e^{2A}\,,\label{eom1}
\end{eqnarray}
together with the first-order Hamiltonian constraint
\begin{equation}
(D-2)(D-3) \dot A^2 + \ft12 (D-3)^2 \dot \phi^2 -\ft1{2} (\lambda_1^2 e^{-a_1\phi} + \lambda_2^2 e^{-a_2 \phi}) e^{2A} = -2(D-2) k^2\,.
\end{equation}
Here a dot denotes a derivative with respect to $\xi$.  We now make a field redefinition
\begin{equation}
\phi=-a_1 q_1 - a_2 q_2\,,\qquad
A=-\ft{(D-3)}{(D-2)}(q_1 + q_2)\,,
\end{equation}
the equations (\ref{eom1}) become
\begin{equation}
\ddot q_1=\lambda_1^2 e^{\fft{4}{N_1} q_1 + \alpha q_2}\,,\qquad
\ddot q_2=\lambda_2^2 e^{\alpha q_1 + \fft{4}{N_2} q_2}\,.\label{eom2}
\end{equation}
where
\begin{equation}
\alpha = a_1 a_2 + \fft{2(D-3)}{D-2}\,.
\end{equation}
For generic $(a_1,a_2)$, the general analytical solution of the Toda-like equations (\ref{eom2}) are unknown.  Special solutions with $\lambda_1\lambda_2=0$, or $a_1\lambda_1^2 + a_2\lambda_2^2=0$ can be constructed.  The former gives rise to singly-charged solutions. For the latter case, we must have $a_1  a_2<0$ and the resulting solution becomes that of Einstein-Maxwell theory.  In each case, the equations are reduced to one Liouville equation, which can be completely solved.

     A more interesting case is that $\alpha=0$, for which,
the Toda-like equations (\ref{eom2}) are reduced to become two independent Liouville equations.  This situation can arise provided that $(a_1,a_2)$ satisfy the condition (\ref{a1a2cons}).  There is the third possibility
\begin{equation}
a_1=-a_2=\fft{6(D-3)}{D-2}\,.
\end{equation}
The resulting equations of motion (\ref{eom2}) become those of the $SL(3,R)$ Toda equations.  The binding energy in this case is not zero but negative in the extremal limit. We shall study this case in another publication \cite{Lu:2013uia}.

    In this paper, we focus on the constraint (\ref{a1a2cons}).
The resulting two independent Liouville equations can be completely solved, giving rise to the non-extremal two-charge black hole
\begin{eqnarray}
ds^2 &=& -(H_1^{N_1} H_2^{N_2}) ^{-\fft{(D-3)}{D-2}} f dt^2 + (H_1^{N_1} H_2^{N_2})^{\fft{1}{D-2}} (f^{-1} dr^2 + r^2 d\Omega_{D-2}^2)\,,\cr
A_1&=& \ft{\sqrt{N_1}\,c_1}{s_1}\, H_1^{-1} dt\,,\qquad
A_2 = \ft{\sqrt{N_2}\,c_2}{s_2}\, H_2^{-1} dt\,,\cr
\phi &=& \ft12 N_1 a_1 \log H_1 + \ft12 N_2 a_2 \log  H_2\,,\qquad
f=1 - \fft{\mu}{r^{D-3}}\,,\cr
H_1&=&1 + \fft{\mu s_1^2}{r^{D-3}}\,,\qquad
H_2 = 1 + \fft{\mu s_2^2}{r^{D-3}}\,.\label{solution1}
\end{eqnarray}
The constraint (\ref{a1a2cons}) implies the following identities
\begin{equation}
N_1 a_1 + N_2 a_2 =0\,,\qquad N_1 + N_2 = \fft{2(D-2)}{D-3}\,.\label{Niident}
\end{equation}
It follows from the second identity in (\ref{Niident}) that both $N_i$ can take integer values only in four and five dimensions, with $N_1 + N_2=3$ and 4 respectively. The solutions with positive integers for $N_i$ are the known black holes in relevant supergravities.

The metric in (\ref{solution1}) approaches Minkowskian space-time at the asymptotic large $r$ region.  One can study
the large-$r$ falloffs and read off the ADM mass \cite{Arnowitt:1959ah} and electric charges, given by
\begin{eqnarray}
M=\ft{(D-2)\omega_{\sst{D-2}}}{16\pi}\mu\Big(1 + \ft{D-3}{D-2}\left(N_1s_1^2 + N_2s_2^2\right)\Big)\,,\qquad
Q_i=\ft{(D-3)\omega_{\sst{D-2}}}{16\pi} \mu \sqrt{N_i} c_i s_i\,,\label{masscharges}
\end{eqnarray}
where $\omega_{\sst{D-2}}$ is the volume of the unit $(D-2)$-sphere.  The outer horizon is located at $r_0=\mu^{1/(D-3)}$, and the temperature and the entropy are given by
\begin{equation}
T= \fft{D-3}{4\pi r_0 c_1^{N_1} c_2^{N_2}}\,,\qquad S=\ft14 c_1^{N_1} c_2^{N_2} r_0^{D-2} \omega_{\sst{D-2}}\,.
\end{equation}
The electric potentials are given by
\begin{equation}
\Phi_i = \sqrt{N_i} \tanh \delta_i\,.
\end{equation}
It is straightforward to verify that the first law of thermodynamics holds, namely
\begin{equation}
dE = T dS + \Phi_1 d Q_1 + \Phi_2 d Q_2\,.\label{firstlaw}
\end{equation}
The inner horizon is located at $r=0$.  There exists an extremal limit in which we send $\mu\rightarrow 0$ while keeping the charges $Q_i$ non-vanishing.  In this limit, the inner and outer horizons coalesce and the near-horizon geometry becomes AdS$_{D-2}\times S^2$.  The mass and the entropy now depend only on the charges, given by
\begin{equation}
M_{\rm ext}=\sqrt{N_1} Q_1 + \sqrt{N_2} Q_2\,,\qquad
S_{\rm ext}=4^{\fft{D-1}{D-3}} ((D-3)\pi)^{\fft{D-2}{D-3}} \left(\ft{Q_1}{\sqrt{N_1}}\right)^{\fft12 N_1}
\left(\ft{Q_2}{\sqrt{N_2}}\right)^{\fft12 N_2}\,.\label{massentropyextr}
\end{equation}
In the extremal limit, the function $f$ becomes 1 and the solution (\ref{solution1}) becomes
\begin{eqnarray}
ds^2 &=& -(H_1^{N_1} H_2^{N_2}) ^{-\fft{(D-3)}{D-2}} dt^2 + (H_1^{N_1} H_2^{N_2})^{\fft{1}{D-2}} dy^m dy^m\,,\cr
A_1&=& N_1\, H_1^{-1} dt\,,\qquad
A_2 = N_2\,c_2\, H_2^{-1} dt\,,\cr
\phi &=& \ft12 N_1 a_1 \log H + \ft12 N_2 a_2 \log  H_2\,,\label{solution2}
\end{eqnarray}
The functions $H_1$ and $H_2$ can now be arbitrary harmonic functions in the $(D-1)$-dimensional flat transverse space $y^m$.  Thus multi-centered or smeared solutions can be obtained.  The extremal RN black hole emerge now in our theory as some harmonic superpositions of the two basic building blocks with appropriate relative charges. The linear summation in the mass formula in (\ref{massentropyextr}) implies that there is no binding energy.
Note that the dilaton coupling constant for Einstein-Maxwell theory is 0, corresponding to $N=2(D-2)/(D-3)$ following the definition of (\ref{Nidef}). Thus the second equation in (\ref{Niident}) is indicative that the RN black hole can arise as the harmonic superposition of the two basic ingredients in our theory.

Thus we see that although the Lagrangian (\ref{lag1}) can always be consistently truncated to Einstein-Maxwell theory provided that $a_1 a_2<0$, for the resulting extremal RN black hole arising as the harmonic superposition of the two more basic ingredients, the dilaton coupling constants must satisfy the more specific condition (\ref{a1a2cons}).  The property of the harmonic superpositions of basic ingredients, which is common for supergravity BPS $p$-branes, is now generalized to be possible in our non-supersymmetric theory in arbitrary dimensions.

   Finally, let us discuss the product formula of entropies of the outer and
inner horizons.  Ignoring the inessential pure numerical constants, the entropies of the outer ($r=r_0)$ and inner $(r=0)$ horizons are given by
\begin{equation}
S_+ \sim \mu^{\fft12 (N_1 + N_2)} c_1^{N_1} c_2^{N_2}\,,\qquad
S_- \sim \mu^{\fft12 (N_1 + N_2)} s_1^{N_1} s_2^{N_2}\,.
\end{equation}
Thus we have
\begin{equation}
S_+ S_- \sim (\mu c_1 s_1)^{N_1} (\mu c_2 s_2)^{N_2} \sim Q_1^{N_1} Q_2^{N_2}\,.
\end{equation}
The fact that the entropy product is expressible in terms of the (quantized) charges only is strongly suggestive of microscopic interpretation of the entropies in terms of some two-dimensional conformal field theories. The entropies can be further expressed as $S_\pm =\sqrt{n_L} \pm \sqrt{n_R}$, where
\begin{equation}
n_L\sim \mu^{N_1+ N_2} (c_1^{N_1}c_2^{N_2} + s_1^{N_1}s_2^{N_2})^2\,,\qquad
n_R\sim \mu^{N_1+ N_2} (c_1^{N_1}c_2^{N_2} - s_1^{N_1}s_2^{N_2})^2\,.
\end{equation}
The $(n_L,n_R)$ are the left and right modes of the two-dimensional conformal field theory, with $n_L-n_R\sim Q_1^{N_1} Q_2^{N_2}$.
These properties were observed in supergravities \cite{CYII,L,CLI,CLII}.  Our results show that it may work also with some rather general non-supersymmetric theories.

\section{Charged asymptotic-AdS black holes}

In the previous section, we found that the Lagrangian (\ref{lag1}) admits analytic charged asymptotic black holes (\ref{solution1}) provided that the dilaton coupling constants satisfy (\ref{a1a2cons}).  In this section, we consider adding a scalar potential with a fixed point so that the solution (\ref{solution1}) generalizes to asymptotic AdS black holes.  We find that the scalar potential is given by
\begin{eqnarray}
V(\phi) &=& -\fft{g^2N_1}{4} \Big[2(D-3)^2 (N_1-1) e^{-a_1\phi} + 2 a_1^2 (D-3)(D-2) N_1 e^{-\fft12(a_1+a_2)\phi}\cr
&&\qquad - a_1^2(D-2)\Big((D-3)N_1 - (D-1)\Big) e^{-a_2\phi}\Big]\,.\label{scalarpot}
\end{eqnarray}
Note that since $(a_1,a_2)$ satisfy the constraint (\ref{a1a2cons}), there is only one non-trivial free parameter $N_1$, in addition to the dimension $D$ and the coupling constant $g$, which should not be confused with the determinant of the metric. The potential has a fixed point at $\phi=0$, with
\begin{equation}
V(0)=-(D-1)(D-2)g^2\,.
\end{equation}
Thus the AdS spacetime with radius $\ell=1/g$ is the vacuum solution of the theory.  Since a massless spin-$s$ field $\psi_s^{M=0}$ in the AdS vacuum satisfies
\cite{Chen:2011in,Lu:2011qx}
\begin{equation}
\Big(\Box + [2-(s-2)(D+s-4)] g^2\Big) \psi_s^{M=0}=0\,,\label{massless}
\end{equation}
we find from the linearized equation of motion of $\phi$ that the scalar is massless, with an asymptotic falloff $1/r^{D-3}$.  Thus the AdS vacuum of our theory does not have any ghost or tachyon instability.

We obtain this scalar potential from the inspiration of those in gauged supergravities and pseudo-supergravities.  In these theories, a black hole solution of the type (\ref{solution1}) of an ungauged theory can be promoted to one of the gauged theory, by simply modifying the function $f$ to include an extra term $g^2 r^2 H_1^{N_1} H_2^{N_2}$, while leaving all other fields unchanged.  We find that the scalar potential (\ref{scalarpot}) can do exactly the same for all the general parameters.  In fact we would obtain the same scalar potential even if we start with the singly-charged black hole and then require that only the function $f$ be modified to become $f=1-\fft{\mu}{r^{D-3}} + g^2 r^2 H_1^{N_1}$ due to the scalar potential.  It is thus rather intriguing that the scalar potential obtained from the field strength $F_1$ with the dilaton coupling constant $a_1$ already ``knows'' the existence of its superposition partner, namely the field strength $F_2$ with $a_2$.

In special cases, the our scalar potential (\ref{scalarpot}) reproduces the ones in gauged supergravities for appropriate parameters.  For example, the $(D,N_1)=(7,2)$ case gives rise to the scalar potential of gauged ${\cal N}=1$ seven-dimensional supergravity \cite{Townsend:1983kk}; the $(D,N_1)=(6,2)$ case gives rise to the scalar potential of ${\cal N}=(1,1)$ six-dimensional gauged supergravity \cite{Romans:1985tw}; the cases with $N_1=1,2$ in five dimensions and $N_1=1,2,3$ in four dimensions can be embedded in maximal gauged supergravities in these dimensions (See e.g.~\cite{tenauthor}).  The theory with only $A_1$ and $N_1=1$ is the bosonic Lagrangian of gauged Kaluza-Klein pseudo-supergravity \cite{Liu:2011ve}, for which the general rotating charged AdS black holes were obtained \cite{Wu:2011zzh}.

Since the potential is inspired by gauged supergravities, it is perhaps not surprising that it can also be expressed in terms of a super potential $W$, namely
\begin{equation}
V=\left(\fft{dW}{d\phi}\right)^2 - \fft{D-1}{2(D-2)} W^2\,,
\end{equation}
with
\begin{equation}
W=\ft{1}{\sqrt2}N_1 (D-3)g\Big(e^{-\fft12a_1 \phi} - \ft{a_1}{a_2} e^{-\fft12 a_2\phi}\Big)\,.\label{suppot}
\end{equation}
General class of domain walls from such a scalar potential was obtained in \cite{lpsdilatonic}. It is of interest to note that the scalar potential of the {\it massive} breathing mode from the sphere reduction of Einstein gravity coupled to an $n$-form field strength also admits a super potential in the form of (\ref{suppot}) but with $a_1 a_2=2(D-1)/(D-2)$ \cite{Bremer:1998zp}.
If we start with a super potential of the type (\ref{suppot}) without restricting the relation between $a_1$ and $a_2$, the resulting scalar potential will give rise to a massive scalar with satisfying $(\Box -m^2)\phi=0$, with
\begin{equation}
m^2 = \ft14\Big((D-2) a_1 a_2 + D-1\Big)^2 -\ft14 (D-1)^2\,,
\end{equation}
where we have chosen the parameter $g$ so that the AdS spacetime has the unit radius.  Thus there is no tachyon instability and the Breitenlohner-Freedman bound is saturated when $a_1 a_2=-(D-1)/(D-2)$.  It follows from (\ref{massless}) that the scalar
becomes massless if $a_1$ and $a_2$ satisfy either the relation (\ref{a1a2cons}) or
\begin{equation}
a_1 a_2 = - \ft{4}{D-2}\,.
\end{equation}
The two possibilities of $a_1a_2$ for a massless scalar become the same in $D=5$.

We now turn our attention to solutions.  We find that the Lagrangian (\ref{lag1}) with the scalar potential (\ref{scalarpot}) and the constraint (\ref{a1a2cons}) admits the static black hole
\begin{eqnarray}
ds^2 &=& -(H_1^{N_1} H_2^{N_2}) ^{-\fft{(D-3)}{D-2}} f dt^2 + (H_1^{N_1} H_2^{N_2})^{\fft{1}{D-2}} (f^{-1} dr^2 + r^2 d\Omega_{D-2}^2)\,,\cr
A_1&=& \sqrt{N_1}\,c_1 s_1^{-1}\, H_1^{-1} dt\,,\qquad
A_2 =\sqrt{N_2}\, c_2 s_2 ^{-1}\, H_2^{-1} dt\,,\cr
\phi &=& \ft12 N_1 a_1 \log H_1 + \ft12 N_2 a_2 \log  H_2\,,\qquad
f=1 - \fft{\mu}{r^{D-3}} + g^2 r^2 H_1^{N_1} H_2^{N_2}\,,\cr
H_1&=&1 + \fft{\mu s_1^2}{r^{D-3}}\,,\qquad
H_2 = 1 + \fft{\mu s_2^2}{r^{D-3}}\,.\label{solution3}
\end{eqnarray}
For appropriate parameters and dimensions, this reproduces some special cases of the charged black holes in gauged supergravities in five \cite{Behrndt:1998jd} and four dimensions \cite{Duff:1999gh}.
The solution with $N_1=1$ in general dimensions was obtained in \cite{Liu:2012ed}.  The solution with only $A_1$ and $N_1=2$ is the special case of the two-charge solutions in \cite{Chow:2011fh} with two charges set equal. The general solution (\ref{solution3}) reduces to the RN-AdS black hole when $\delta_1=\delta_2$.  Note that the solution for the scalar is independent of $g$.  This is a consequence that the scalar is massless and hence its falloff $1/r^{D-3}$ is dependent of $g$.

The asymptotic region of (\ref{solution3}) at $r=\infty$ is the AdS boundary in global coordinates.  The mass and charges are still given by (\ref{masscharges}) and they are independent of the charge parameters.  We followed the procedure of \cite{Chen:2005zj} to calculate the mass, which was based on the definition of conformal mass \cite{Ashtekar:1984zz,Ashtekar:1999jx}. The outer horizon is located at the largest real root $r_0$ of $f(r)=0$.  The temperature, entropy and electric potentials are then given by
\begin{eqnarray}
T&=&\fft{f'(r_0)}{4\pi H_1(r_0)^{N_1/2} H_2(r_0)^{N_2/2}}\,,\qquad S=\ft14 r_0^{D-2} H_1(r_0)^{N_1/2} H_2(r_0)^{N_2/2} \Omega_{D-2}\,,\cr
\Phi_i &=& \sqrt{N_i}\, \fft{c_i}{s_i} \Big(1 - \fft1{H_i(r_0)}\Big)\,.
\end{eqnarray}
It is straightforward to verify that the first law of thermodynamics (\ref{firstlaw}) continue to hold.

It is worth commenting that the $r=0$ surface is no longer one of the inner horizons, since $g_{tt}$ at $r=0$ becomes non-vanishing.  It follows that the ``extremal limit'' discussed above (\ref{massentropyextr}) now gives rise to a solution with naked singularity. In gauged supergravities, such solutions are supersymmetric and are called ``superstars.''  The supersymmetric of such solution was first analysed in \cite{Romans:1991nq}
in the context of Einstein-Maxwell theory with a cosmological constant.  (The generalization to the rotating case was given in \cite{AlonsoAlberca:2000cs}.) The resolution of the singularity in supergravities while maintaining some supersymmetry is either through the bubbling AdS procedure \cite{Lin:2004nb} and turning the superstars to smooth solitons \cite{Lin:2004nb,Chong:2004ce} or by adding rotation and constructing supersymmetric rotating black holes \cite{Gutowski:2004ez}.   (In four dimensions, the BPS magnetic $(D-4)$-branes discussed in section 6 become magnetic black holes.  Such solutions \cite{Cacciatori:2009iz} and their non-extremal generalization \cite{Toldo:2012ec,Klemm:2012vm} were previously obtained in four-dimensional gauged supergravity.) Whether such resolutions also exist in our general non-supersymmetric theory is far from clear, since charged rotating black holes are notoriously difficult to construct in general dimensions and the soliton approach requires additional axionic scalars \cite{Chong:2004ce}.

   There exists alternative extremal limit in which the function $f$
acquires a double zero for the largest root, namely $f(r_0)=0=f'(r_0)$, but with $f''(r_0)\ne 0$.  For this choice of parameters, the near-horizon geometry is the AdS$_2\times S^{D-2}$.  There is a further intriguing limit when one of the charge parameter is set to zero, say $\delta_2=0$. The function $f$ is now given by
\begin{equation}
f=1 - \fft{\mu}{r^{D-3}} + g^2 r^2 H_1^{N_1}\,.
\end{equation}
For $N_1=\fft{D-1}{D-3}$, if we let
\begin{equation}
\mu = g^2 (\mu s_1^2)^{N_1}\,,
\end{equation}
the $1/r^{D-3}$ pole of $f$ at $r=0$ cancels and $f$ becomes a non-vanishing constant as $r\rightarrow 0$.  The $r=0$ is then a null horizon where the horizon and curvature singularity coincide.  The near-horizon geometry at $r=0$ takes the same structure as the extremal solution (\ref{solution2}) that is asymptotic to the flat space-time.  The near-horizon geometry becomes AdS$_2\times S^{D-2}$ with a hyper scaling.  To be specific, the metric $e^{\phi/a_1} ds^2$ becomes AdS$_2\times S^{D-2}$ around $r=0$.  The $(N_1,D)=(2,5)$ and $(3,4)$ solutions can be embedded in five and four-dimensional gauged supergravities.  As we shall see in the end of this section, a simple scaling can turn the $S^{D-2}$ to $R^{D-2}$ or $H^{D-2}$.  Thus AdS$_2\times R^{D-2}$ (or AdS$_2\times H^{D-2}$) with a hyperscaling can also arise as the near-horizon geometry in our solutions. The five-dimensional solution was studied in \cite{Gubser:2012yb}.  A general discussion on the infrared AdS$_2\times R^2$ can be found in \cite{Bhattacharya:2012zu}.

The entropy product formulae for the AdS charged black holes are more complicated.  Although $r=0$ is no longer one of the inner horizon, there are larger number of roots for $f=0$.  It was shown using many explicit examples in gauged supergravities that the entropy product formulae in terms of quantized charges can still be true if one consider the
``entropies'' for the horizons associated with all the roots of $f$ \cite{cvgipo}.  In our general solutions, if $N_i$ are rational numbers, then $f=0$ can be turned into a polynomial equations of some finite order, and hence there is a finite number of horizons.  we expect that the entropy product formula is
\begin{equation}
\prod_{i=1}^{2(D-2)} S_i \sim \Big[(g^{-1} Q_1)^{N_1} (g^{-1}Q_2)^{N_2}\Big]^{D-3}\,.\label{entropyprod}
\end{equation}
We are unable to give a full proof.  The formula does reproduce all the correct answer when our theory can indeed embedded in supergravities.  It also reproduces the result for the general RN-AdS black holes. Furthermore, we have checked many explicit examples outside supergravities, and (\ref{entropyprod}) always holds.  Let us present an explicit example: $D=7$ and $N_1=1$ and hence $N_2=3/2$.  Defining $\rho=r^4$, we find that the equation $f=0$ implies the vanishing of a quintic polynomial:
\begin{eqnarray}
&&\rho^5+(2q_1 + 3q_2 - g^{-4}) \rho^4 + (q_1^2 + 6q_1 q_2 + 3q_2^2 + 2\mu g^{-4}) \rho^3\cr
&&\qquad + (q_2(3q_1^2 + 6q_1 q_2 + q_2^2) - \mu^2 g^{-4}) \rho^2 +
q_1 q_2^2 (3q_1 + 2q_2) \rho + q_1^2 q_2^3=0\,,\label{rhoeom}
\end{eqnarray}
where $q_i=\mu \sinh^2\delta_i$. Thus, there are five roots for $\rho$ and hence a total of 20 roots for the $r$ coordinate. Since the equation of (\ref{rhoeom}) involves $g^4$ whilst the function of $f$ involves only $g^2$,  only ten roots of $r$ are related to the null surfaces.   The entropy on each horizon associated with $\rho_i$ is given by
\begin{equation}
S_i\sim (\rho_i + q_2)^{\fft12} (\rho_i + q_2)^{\fft32}\,,\qquad i=1,2,\cdots,5\,.
\end{equation}
It is then straightforward to verify that (\ref{entropyprod}) is indeed valid.

Finally for static AdS black holes, one can also construct solutions whose level surfaces are tori or hyperbolic spaces. The solutions with torus topology are particularly important since the asymptotic AdS space-time is then in the Poincar\'e patch, which is particularly suitable for the discussion of the AdS/CFT correspondence.  These topological black holes can be easily obtained by some appropriate scaling.  Let
\begin{equation}
r^2\rightarrow \epsilon^{-1}\, r^2\,,\qquad t\rightarrow \epsilon^{\fft12}\, t\,,\qquad \delta_i\rightarrow \epsilon^{\fft12}\,\delta_i\,,\qquad
d\Omega_{D-2}^2 \rightarrow \epsilon\, d\Omega_{D-2,\epsilon}^2\,,
\end{equation}
together with some appropriate scaling of $\mu$.  It is easy to verify that the limit of $\epsilon=0$ is smooth and that of $\epsilon=-1$ is real. The function $f$ for the resulting topological black holes take the form
\begin{equation}
f=\epsilon - \fft{\mu}{r^{D-3}} + g^2 r^2 H_1^{N_1} H_2^{N_2}\,,
\end{equation}
where $\epsilon=1,0,-1$. The rest of the solutions can be obtained easily and we shall not present here to avoid repetitions.

\section{$D=4$ dyonic black hole and the lift to $D=11$}

In four dimensions, for $N_1=2$, we have $a_1=-a_2=1$, the equations of motion of the two-charge black hole associated with two different field strength are identical to those of the dyonic black hole with both the electric and magnetic charges carried by the same field strength.  The Lagrangian is given by
\begin{equation}
e^{-1}{\cal L}_4 = R - \ft12(\partial\phi)^2 - \ft14 e^{\phi} F^2 + 2 (2 + \cosh\phi)g^2\,.
\label{d4lag}
\end{equation}
This Lagrangian can be consistently embedded in gauged ${\cal N}=4$ $SO(4)$ gauged supergravity \cite{Das:1977pu}, for which the explicit $S^7$ reduction ansatz from $D=11$ was obtained in \cite{Cvetic:1999au}.  Following the previous discussion, the Lagrangian
(\ref{d4lag}) admits the following dyonic solution\footnote{In the $U(1)^4$ solution of four-dimensional ungauged supergravity, each individual field strength are either electric or magnetic, but not both from the point of view of the string or M-theory.  Nevertheless the solution has been called dyonic because if one lifts the solutions to higher dimensions, the four ingredients cannot be all electric or magnetic.  The $U(1)^4$ black hole in gauged supergravity however are all electric or magnetic.  In this paper,a solution is dyonic if and only if the same field strength carries both electric and magnetic charges. Another known example is the Kaluza-Klein dyonic black hole. (See, e.g.~\cite{Gibbons:1985ac}.)}
\begin{eqnarray}
ds^2 &=& -\fft{f}{H_1H_2} dt^2 + H_1 H_2 \Big(\fft{dr^2}{f} + r^2 d\Omega_2^2\Big)\,,\cr
A &=& \sqrt2\, \fft{c_1}{s_1} H_1^{-1} dt + \sqrt{2}\, \mu c_2 s_2 \omega\,,\qquad
\phi = \log\fft{H_1}{H_2}\,,\cr
f&=& 1 - \fft{\mu}{r} + g^2 r^2 H_1^2 H_2^2 \,,\qquad H_i = 1 + \fft{\mu s_i^2}{r}\,,
\label{d4solution1}
\end{eqnarray}
where $\omega$ is the 1-form with $d\omega=\Omega_\2$, the volume form of the unit 2-sphere metric $d\Omega_2^2$. We now study the higher-dimensional origin of this solution.  Using the reduction ansatz \cite{Cvetic:1999au}, we find that the eleven-dimensional solution is
\begin{eqnarray}
ds_{11}^2 &=&\Delta^{\fft23}\Big[ - (H_1H_2)^{-1} f dt^2 +  H_1H_2 (f^{-1} dr^2 + r^2 d\Omega_2^2) + 4 g^{-2} d\xi^2 \cr
&&+ g^{-2}\Big(\fft{c^2 H_1^2}{c^2 H_2^2 + s^2} \Big(\sigma_1^2 + \sigma_2^2 + (\sigma_3 + c_1s_1^{-1} H_1^{-1} dt + \mu c_2 s_2 \omega)^2\Big)\cr
&&\qquad\qquad +
\fft{c^2 H_2^2}{c^2 H_1^2+ s^2} d\Omega_3^2\Big)\Big]\,,\cr
F&=& -g (H_1^{-2} H_2^{2} c^2 + H_1^2 H_2^{-2} s^2 + 2) \epsilon_4 - 4g^{-1} sc H_1 H_2^{-1} {*_4 d} (H_1/H_2) \wedge d\xi\,,
\end{eqnarray}
where $c=\cos\xi$ and $s=\sin\xi$ and $\epsilon_4$ is the volume 4-form of the metric in (\ref{d4solution1}). The notation $*_4$ denotes the four-dimensional Hodge dual and $\sigma_i$ are the $SU(2)$ left-invariant 1-forms, satisfying $d\sigma_i=\fft12 \epsilon_{ijk} \sigma_j\wedge \sigma_k$.  It is clear that the electric component associated with $H_1$ describes a rotation in the space-time.  The magnetic component associated with $H_2$ describes an $S^7$ bundle over $S^2$. Had we considered the four-dimensional solution carried by two field strengths instead, the eleven-dimensional solution will drop the bundle structure and add another rotation in the $d\Omega_3^2$ direction.

Let us now consider turning off the scalar potential by setting $g=0$.  The theory can then be embedded in ungagued supergravity.  The solution with the electric charge can be uplifted to become self-dual string in $D=6$, which itself can be lifted to becomes the D1/D5 (or M2/M5) intersections, with equal D1 and D5 (or M2 and M5) charges.  The same is true with the pure magnetic black hole.  Thus the dyonic solution describes some non-supersymmetric intersection of two self-dual strings.  This implies that the D1/D5 system with equal charges can intersect with another such D1/D5 on to a dyonic black hole.  The same story applies to the M2/M5 system.  In the extremal limit, these intersections, although non-supersymmetric, are also characterized by the harmonic superpositions.  To demonstrate this explicitly, let us follow the discussion in section 2, and take the extremal limit of (\ref{d4solution1}), we find
\begin{eqnarray}
ds^2_4 &=& (H_1H_2)^{-1} dt^2 + H_1 H_2 dy^m dy^m\,,\qquad
\phi=\log\fft{H_1}{H_2}\,,\cr
F &=& \sqrt2\, dH_1^{-1}\wedge dt + \sqrt2\, (H_1H_2)^{-1}  {* (dH_2\wedge dt)}\,,
\end{eqnarray}
where $H_1$ and $H_2$ are arbitrary harmonic functions in the Euclidean 3-space $y^m$.  Assuming that the four-dimensional field strength $F_2$ in (\ref{d4lag}) arises from the reduction of the field strengths in $D=11$ or $D=10$, rather than from the metric, we summarize one possibility of the lifting in table \ref{intersection}.
\bigskip

\begin{table}[ht]
\begin{center}
\begin{tabular}{|c|cccc|c c|c c c c|c|}
  \hline
 & t & $y_1$ & $y_2$ &$y_3$ & $x_1$ & $x_2$ & $z_1$ & $z_2$ & $z_3$ &$z_4$ & $z_5$\\ \hline
 $H_1$ & x & -- & -- & -- & -- & x & -- & -- & -- & -- & x \\
       & x & -- & -- & -- & -- & x & x  & x  & x  & x  & -- \\ \hline
 $H_2$ & x & -- & -- & -- & x & -- & -- & -- & -- & -- & x \\
       & x & -- & -- & -- & x & -- & x  & x  & x  & x  & -- \\ \hline
\end{tabular}
\end{center}
\caption{\footnotesize The harmonic intersecting rules of the M2/M5-branes (or D1/D5 or self-dual string.) The $H_1$ is the harmonic function for one M2/M5 system with equal M2 and M5 changes and the $H_2$ is for another.}\label{intersection}
\end{table}

Let us now explain in some detail the intersection rule in table \ref{intersection}. It should be emphasized that the electric black hole associated with the harmonic function $H_1$ is not itself a fundamental building block in string or M-theory, but it is a threshold bound state of two more fundamental states with {\it equal} charges so that the intersection is described by only one harmonic function. The same story applies to the magnetic solution associated with $H_2$.  This is the reason why each $H_i$ in table \ref{intersection} is associated with two $p$-branes.  In the table, ``x'' denotes the world-volume direction of a $p$-brane whilst the symbol ``--'' denotes the transverse space.  The coordinates $(t,y_1,y_2,y_3)$ describe the four-dimensional space-time of the dyonic black hole. If we lift the solution to five-dimensions to include $x_1$, the electric solution (associated with $H_1$) remains a black hole whilst the magnetic solution ($H_2$) becomes a string.  Lift the solution further to include $x_2$, both electric and magnetic solutions become self-dual strings, carried by the same self-dual 3-form field strength in six dimensions.  If we lift the solution further to type IIB theory with the internal 4-space $(z_1,z_2,z_3,z_4)$, and let the solution be supported by the R-R 3-form, the solution then describes the intersection of two D1/D5-systems.  Lifting the solution back to $D=11$ giving rise to the intersection of two M2/M5-branes.
To conclude this discussion, let us present the explicit solution of the (M2/M5)-(M2/M5) intersection
\begin{eqnarray}
ds_{11}^2 &=& -(H_1H_2)^{-1} dt^2 + (H_1 H_2) dy^m dy^m + H_1 H_2^{-1} dx_1^2 + H_1^{-1} H_2 dx_2^2 + dz_i dz_i\,,\cr
F_4 &=& \sqrt2 \Big[dH_1^{-1} \wedge dt \wedge dx_2\wedge dz_5 + \ft1{2} \epsilon^{m}{}_{nk} \partial_m H_1\, dy^n\wedge dy^k\wedge dx_1 \wedge dz_5\cr
&&\qquad+ dH_2^{-1} \wedge dt \wedge dx_1\wedge dz_5 + \ft1{2} \epsilon^{m}{}_{nk} \partial_m H_2\, dy^n\wedge dy^k\wedge dx_2 \wedge dz_5\Big]\,.
\end{eqnarray}
Other intersections, such as D1/D5 with itself or intersecting self-dual strings and the non-extremal versions, can all be derived straightforwardly, and we shall not present them here.  The 4-space $(z_1,z_2,z_3,z_4)$ can be also the K3 manifold.

\section{Magnetic AdS$_{D-2}\times {\cal Y}^2$ vacua}

We now consider the solutions with the vectors $A_i$ carrying the magnetic charges.  If we turn off the scalar potential by setting $g=0$, the Lagrangian (\ref{lag1}) with (\ref{a1a2cons}) admits the two-charge $(D-4)$-branes:
\begin{eqnarray}
ds^2_D &=& (H_1^{N_1} H_2^{N_2})^{-\fft{1}{D-2}} (-f dt^2 + dx^i dx^i) + (H_1^{N_1} H_2^{N_2})^{\fft{D-3}{D-2}} (f^{-1} dr^2 + r^2 d\Omega_2^2)\,,\cr
A_i &=& \sqrt{N_i} s_i c_i\, \omega\,,\qquad \phi = -\ft12 N_1 a_1 \log H_1 -\ft12 N_2 a_2 \log H_2\,,\cr
f&=& 1- \fft{\mu}{r}\,,\qquad H_i = 1 + \fft{\mu s_i^2}{r}\,.
\end{eqnarray}
In the extremal limit, the solution interpolates between the AdS$_{D-2}\times S^2$ in the horizon to the asymptotic $D$-dimensional Minkowski spacetime.  These solutions are magnetic duals to the electrically-charged black holes.

When the scalar potential (\ref{scalarpot}) is included, the magnetic $(D-4)$-brane no longer takes the analogous form except in $D=4$. In four dimensions, the electric field strengths $F_i$ can be dualized to carry magnetic charges, together with $\phi\rightarrow -\phi$.  If we choose parameters such that the function $f$ has a double largest root, the solution has a decoupling limit of magnetic AdS$_2\times {\cal Y}^2$ where ${\cal Y}^2$ can be $R^2, S^2$ or hyperbolic 2-space.  However, when the theory can be embedded in (maximal) gauged supergravity in four dimensions, such a magnetically-charged black hole cannot be supersymmetric \cite{Duff:1999gh}, and hence neither the magnetic AdS$_2\times {\cal Y}^2$ vacua constructed this way.

In this section, we are interested in constructing vacuum solutions AdS$_{D-2}\times {\cal Y}^2$ in our theory in general dimensions.  Furthermore, we would like the solutions to be supersymmetric when the theory can indeed be embedded in gauged supergravities.  Let us first examine the full bosonic equations.  The ansatz is given by
\begin{equation}
ds_{D}^2 = ds_{\sst{AdS}}^2 + \ell^2\, d\Omega_{2,\epsilon}^2\,, \qquad F_i = q_i \ell^2 \Omega_{\2}\,,
\end{equation}
where the metrics $ds_{\sst{AdS}}^2$ and $d\Omega_{2,\epsilon}^2$ have the following Ricci curvature
\begin{equation}
R_{\mu\nu} = -\ft{D-1}{L^2}\, g_{\mu\nu}\,,\qquad R_{ij} = \fft{\epsilon}{\ell^2} g_{ij}\,,
\end{equation}
with $\epsilon=1,0,-1$, respectively, and $\Omega_\2$ is the volume 2-form for $d\Omega_{2,\epsilon}^2$.  The dilaton is taken to be constant.  The equations of motion from the Lagrangian (\ref{lag1}) with the potential (\ref{scalarpot}) and the constraint (\ref{a1a2cons}) can be reduced to
\begin{eqnarray}
&&q_1^2 e^{a_1 \phi} + q_2^2 e^{a_2 \phi} = \ft{2(D-1)}{L^2} + \fft{2\epsilon}{\ell^2}\,,\cr
&&a_1 q_1^2 e^{a_1 \phi} + a_2 q_2^2 e^{a_2 \phi} = 2\ft{dV}{d\phi}\,,\cr
&&V= -\ft{(D-1)(D-3)}{L^2} + \fft{\epsilon}{\ell^2}\,.\label{adsy2}
\end{eqnarray}
There are a total of five unknown variables $(L, \ell, q_1, q_2, \phi)$ and three equations, thus solutions exist in general.  Of course, for the solution to be physical, these quantities all have to be real.

    When the theory can be embedded in supergravities, some of the
vacuum solutions (\ref{adsy2}) can be supersymmetric. A sure way to discover supersymmetric solutions is to examine the supersymmetric transformation rules of the fermions.  This is however theory dependent and there is no general approach.  An alternative but not always reliable way is to use the super potential method \cite{Cucu:2003bm,Cucu:2003yk}.  This method is particularly applicable in our case since we have not yet established whether consistent Killing spinor equations exist in our general theory.  Following the procedure outlined in \cite{Cucu:2003bm,Cucu:2003yk}, we first perform the dimensional reduction on ${\cal Y}^2$ upon the Lagrangian (\ref{lag1}) with the scalar potential (\ref{scalarpot}) with the reduction ansatz
\begin{eqnarray}
ds_D^2 &=& e^{2\alpha \varphi} ds_{D-2}^2 + e^{2\beta\varphi} \ell^2 d\Omega^2_{2,\epsilon}\,,\qquad F_i = q_i \ell^2 \Omega_\2\,,\cr
\alpha &=& -\ft{1}{\sqrt{(D-2)(D-4)}}\,,\qquad \beta = - \ft12 (D-4)\alpha\,.
\end{eqnarray}
The $(D-2)$-Lagrangian is given by
\begin{eqnarray}
e^{-1} {\cal L} &=& R - \ft12 (\partial\phi)^2 - \ft12 (\partial \varphi)^2 - \widetilde V\,,\cr
\widetilde V &=& \ft12 (q_1^2 e^{a_1\phi} + q_2^2 e^{a_2\phi}) e^{2(D-3)\alpha \varphi} - 2\epsilon \ell^{-2} e^{(D-2)\alpha \varphi} + e^{2\alpha\varphi} V\,.
\end{eqnarray}
We would like to find a super potential $\widetilde W$ such that
\begin{equation}
\widetilde V=(\ft{\partial \widetilde W}{\partial \phi})^2 + (\ft{\partial \widetilde W}{\partial \varphi})^2 - \ft{D-3}{2(D-4)} \widetilde W^2\,.
\end{equation}
The situation with $\epsilon=0$ is different from $\epsilon=\pm1$ and hence we shall discuss them separately.

\bigskip
\noindent{\bf Case 1: $\epsilon=0$}
\medskip

Let us first consider $\epsilon=0$, corresponding to ${\cal Y}^2$ being $R^2$.  Such solutions in gauged supergravities were implicitly obtained in \cite{Cucu:2003bm,Cucu:2003yk}, in that the scalar curvature of the ${\cal Y}^2$ was shown in \cite{Cucu:2003bm,Cucu:2003yk} to be proportional to the sum of all the charges $\tilde q_1 + \tilde q_2 + \cdots$.  For our theory, we find that there exists a super potential
\begin{equation}
\widetilde W=\sqrt{\ft12 N_1} \, q_1 (e^{\fft12 a_1\phi} - e^{\fft12 a_2\phi}) e^{\alpha (D-3)\varphi} + e^{\alpha \varphi} W\,,
\end{equation}
where $W$ is given by (\ref{suppot}), provided that the charges $q_i$ satisfy the following constraint
\begin{equation}
N_1 q_1^2 = N_2 q_2^2\,.\label{epsilon0cons}
\end{equation}
Thus the existence of such a super potential requires turning on both field strengths, in which case, only the five and four dimensional solutions of our theory have the possibility to become those of supergravities.  In fact, supersymmetric AdS$_3\times R^2$ vacua was obtained using Killing spinor analysis in the $U(1)^3$ gauged supergravity where the three charges satisfy the condition $\tilde q_1 + \tilde q_2 + \tilde q_3=0$ \cite{Almuhairi:2010rb,Almuhairi:2011ws}.  If one sets two charges equal, the condition becomes equivalent to ours.  To see this, let $\tilde q_1=q_1$ and $\tilde q_2=\tilde q_3=q_2/\sqrt2$, we obtain (\ref{epsilon0cons}).  Our procedure will not be able to produce the AdS$_5\times R^2$ solution in $D=7$ gauge supergravities since the embedding requires either $N_1=1$ and $F_2=0$ or $N_2=2$ and $F_2=0$.  The former corresponds to the single charge and the latter to two-equal charges, whilst the AdS$_5\times R^2$ emerges with $\tilde q_1+\tilde q_2=0$.  The $D=7$ solutions obtained in \cite{Cucu:2003bm,Cucu:2003yk} using the super potential method were independently obtained using Killing spinor analysis. (See,~e.g.~\cite{Anderson:2011cz}.)

The ``supersymmetric'' subset of the fix points satisfy
\begin{equation}
\ft{\partial \widetilde W}{\partial \phi} = 0 = \ft{\partial \widetilde W}{\partial \varphi}\,,
\end{equation}
which are guaranteed to be also the fixed points of $\widetilde V$.  We find the following two fix points for a given set of parameters
\begin{eqnarray}
e^{\fft{2}{a_1 N_1} \phi} &=& \fft{(D-3)((D-2)a_1 N_1 \pm 2\sqrt{2(D-2)})}{a_1 (D-2) ((D-3) N_1 - 2)}\equiv \gamma\,,\cr
e^{2\beta\varphi} &=& \fft{(D-3) \sqrt{N_1}\, q_1 (\gamma-1)\gamma^{\fft{(D-2) - N_1(D-3)}{D-2}} }{g((D-3)N_1 (\gamma-1) - 2(D-2) \gamma}\,.
\end{eqnarray}
We can read off the cosmological constant $\widetilde V$; however, the expression is very complicated for general cases.  Let us consider some concrete examples.  The $N_1=2$ case is relatively simpler.  There is only one ``supersymmetric'' fixed point, given by
\begin{eqnarray}
&&e^{\sqrt{(D-2)/2} \phi} = \ft{2(D-3)}{D-4}\,,\qquad
e^{2\beta \varphi} = \ft{-q_1}{\sqrt2\,g} \, (\ft{D-4}{2(D-3)})^{\fft{D-4}{D-2}}\,,\cr
&&\widetilde V = - \ft{(D-2)^2(D-3) g^2}{D-4} (\ft{\sqrt2 g}{-q_1})^{\fft2{D-4}}\,.
\end{eqnarray}
Lift the AdS$_{D-2}$ vacuum back to $D$-dimensions, we obtain the AdS$_{D-2}\times R^2$ solution
\begin{eqnarray}
ds_D^2 &=& L^2 (d\rho^2 + e^{2\rho} dx^\mu dx^\nu\eta_{\mu\nu} + dz_1^2 + dz_2^2)\,,\cr
F_1 &=& \ft{2\sqrt2 (D-3)(D-4)}{(D-2)^2g}  dz_1\wedge dz_2\,,\qquad
F_2 = \ft{2\sqrt{2(D-1)(D-3)} (D-4)}{(D-2)^2g} dz_1\wedge dz_2\,,\cr
L&=&\ft{D-4}{g(D-2)} (\ft{2(D-3)}{D-4})^{1/(D-2)}\,,\qquad e^{\sqrt{\fft12(D-2)}\, \phi} = \ft{2(D-3)}{D-4}\,.\label{adsr2}
\end{eqnarray}
Note that the parameter $q_1$ drops out in the $D$-dimensional solution.  This AdS$_{D-2}\times R^2$ vacuum (\ref{adsr2}) may be viewed as the near-horizon geometry of some extremal magnetic AdS black $(D-4)$-brane.  To study the stability of the solution (\ref{adsr2}), we can examine the masses of the linear fluctuations of the scalar modes $\phi$ and $\varphi$ around the AdS$_{D-2}\times T^2$ vacua.  We find that the eigenvalues of the quadratic mass matrix are given by
\begin{equation}
m_\pm^2 = \ft{(D-3)^2}{(D-2)^2} \Big( 3D^2 - 24 D + 52 \pm (D-6) \sqrt{(D-2)(9D-34)}\Big)\,.
\end{equation}
Thus there is no instability associated with a tachyon mode.  This is of course ensured by our super potential construction.

For $N_1=1$, the results are more complicated.  In $D=5$, it is equivalent to $N_1=2$ and we shall not present here.  We shall only present $D=6$ and $D=7$ examples:
\begin{eqnarray}
D=6:&& e^{2\sqrt{2/5}\,\phi} = \ft15 (15\pm 6\sqrt5) \,,\qquad
e^{\varphi/\sqrt2}=\mp (75\pm \sqrt5)^{\fft14} \ft{q_1}{5g}\,,\cr
&&\qquad \widetilde V=\pm 20\sqrt{\ft23(5\mp \sqrt5)}\, q_1 g^3\,,\cr
D=7:&& e^{\sqrt{5/3}\phi} = 2\pm 2\sqrt{\ft23}\,,\quad e^{\sqrt{3/5}\,\varphi} = \mp \ft{(\sqrt6 \pm 2)^{1/5}}{108^{1/5}} q_1 g^{-1}\,,\cr
&&\qquad\widetilde V=-g^2(131 \mp 16 \sqrt{6})^{1/3} g^2 (\mp g q_1^{-1})^{2/3}\,.
\end{eqnarray}
Owing to the increasing complexity we shall not present any explicit example further.

\bigskip
\noindent{\bf Case II: $\epsilon=\pm 1$}
\medskip

This was well studied in gauged supergravities \cite{Cucu:2003bm,Cucu:2003yk}.
Let us first turn on only $q_1$ by setting $q_2=0$.  We find that the existence of the super potential requires that
\begin{equation}
q_1 = \fft{2\epsilon}{(D-3)\sqrt{N_1}\,g\ell^2}\,.
\end{equation}
The super potential is then given by
\begin{equation}
\widetilde W=\ft{\sqrt2\, \epsilon}{(D-3)g\ell^2} e^{\fft12 a_1 \phi + (D-3)\alpha \varphi} + e^{\alpha \varphi} W\,.
\end{equation}
The detail discussion for $N_1=1,2$ and $D\le 7$ was given in \cite{Cucu:2003bm,Cucu:2003yk} .  In general, we have
\begin{equation}
e^{\fft{2}{a_1 N_1}\phi} = \ft{(D-3)N_1}{(D-3)N_1 -2}\,,\qquad
e^{2\beta\varphi} = \fft{-\epsilon}{(D-3)g^2\ell^2} \left(\ft{(D-3)N_1}{(D-3)N_1 - 2}\right)^{\fft{(D-3)N_1 - (D-2)}{D-2}}\,,
\end{equation}
Thus we must have $\epsilon=-1$ in this case, and the ${\cal Y}^2$ has to be a hyperbolic 2-space.  The potential at the fixed point is
\begin{equation}
\widetilde V=-(D-4)g^2 (D-3)^{1 + \fft{(D-3)N_1}{D-4}} ((D-3)N_1 -2)^{\fft{2-(D-3)N_1}{D-4}} (g\ell)^{\fft{4}{D-4}}\,.
\end{equation}
The negativity of the potential implies that the $D$-dimensional space-time is the AdS$_{D-2}\times H^2$.  The supersymmetric solution of $N_1=1$ in seven-dimensional gauged supergravity was obtained in \cite{Maldacena:2000mw}. (See also, \cite{Cacciatori:1999rp,Klemm:2000nj}.)

Finally let us consider $q_1q_2\ne 0$.  In this case, the condition for the existence of the super potential is given by
\begin{equation}
\sqrt{N_1}\, q_1 + \sqrt{N_2}\, q_2 = \fft{2\epsilon}{(D-3) g\ell^2}\,.\label{q1q2cons}
\end{equation}
The super potential is given by
\begin{equation}
\widetilde W= e^{\alpha \varphi} W +\ft1{\sqrt2} e^{(D-3)\alpha \varphi} \Big(\sqrt{N_1}\, q_1 e^{\fft12 a_1\phi} + \sqrt{N_2}\, q_2 e^{\fft12 a_2\phi}\Big)\,.
\end{equation}
As in gauged supergravities \cite{Cucu:2003bm,Cucu:2003yk}, we find that AdS$_{D-2}\times S^2$ solutions, with $\epsilon=+1$,  can also arise in general dimensions.  To present the solutions, it is perhaps more convenient to solve for $q_1/g$ and $g^2\ell^2$ in terms of constants $(\phi,\varphi)$, we find
\begin{eqnarray}
q_1g^{-1} &=& \ft12(d-3)\sqrt{N_1} e^{2\beta\varphi} \Big((2-N_1) e^{-a_1\phi} - N_2 e^{-\fft12(a_1+a_2)\phi}\Big)\,,\cr
(g\ell)^{-2} &=& \ft14 e^{2\beta \varphi} \Big(N_1(2-N_1)(D-3)^2 e^{-a_1\phi} - 2 N_1 N_2 (D-3)^2 e^{-\fft12(a_1+a_2)\phi}\cr
&&\qquad\quad +
(D-3)((D-3)N_1 -2)N_2 e^{-a_2\phi}\Big)\,.
\end{eqnarray}
Note that the reality condition of $g\ell$ imposes a condition on the allowed range of the scalar $\phi$.  General class of AdS$_{D-2}\times S^2$ solutions in gauged supergravities were giving in \cite{Cucu:2003bm,Cucu:2003yk}.  The $D=7$ example provides a smooth embedding of AdS$_5$ in M-theory \cite{Cucu:2003bm,Cucu:2003yk}.

Finally, it is worth commenting that the first-order equations can also be easily derived from the super potential, providing flows from the infrared AdS$_{D-2}\times {\cal Y}^2$ to the UV AdS$_{D}$, as was shown in \cite{Cucu:2003bm,Cucu:2003yk}.

\section{Conclusions}

In this paper we have constructed a bosonic theory (\ref{lag1}) with the scalar potential (\ref{scalarpot}) and the constraint (\ref{a1a2cons}). The theory is the simplest generalization of the phenomena that Einstein-Maxwell gravities in four and five dimensions can be embedded in the $U(1)^3$ and $U(1)^4$ theories of (gauged) supergravities.  The generalizations are two folds: one is that our theory can be in general dimensions; the other is that the dilaton coupling constants $a_i$ can take any real number subjecting to the constraint (\ref{a1a2cons}).

One reason that we believe such generalization should exist is the progress in understanding pseudo-supergravities where the solutions share many same properties of those in true (on-shell) supergravities. In the case with no scalar potential, the extremal RN black hole in any dimension emerges as the bound state in our theory of the basic building blocks carrying $F_1$ and $F_2$ charges, with zero binding energy.  The entropy product rule of the non-extremal black holes is also suggestive of microscopic interpretation of the entropies in terms of some two-dimensional conformal field theories, as in the case of the examples in supergravities.

When the scalar potential is turned on, charged AdS black holes can be obtained in the similar fashion as those in gauged supergravities.  We analyse the global structure and thermodynamics and obtain that the product of the entropies of all horizons is expressed in terms of charges only.  We can also construct the AdS$_{D-2}\times {\cal Y}^2$ vacua, which may arise as the near-horizon geometries of some $(D-4)$-branes, the magnetic duals to the electrically-charged AdS black holes.  We concentrate on the ``supersymmetric'' vacua that arise as the fixed points of some super potentials.  They indeed recover the known supersymmetric vacua derived from the Killing spinor equations when the theory can be embedded in supergravities.   We also verified using some examples of the AdS$_{D-2}\times R^2$ that tachyon instability is absent in these vacua, a natural consequence of the super potential method.  These properties are suggestive that our theory can be embedded in some pseudo-supergravities. The fact that the scalar turns out to be massless in our theory is consistent with this conjecture.  Indeed, the pseudo-supergravity extension of our theory with only $F_1$ and $N_1=1$ was already constructed without needing any additional bosonic field \cite{Liu:2011ve}.

As a bonus, we also find a new type of intersections where the M2/M5-brane system with equal charges can be harmonically superposed with another such M2/M5 to give rise to a dyonic black hole.  The same story applies to the D1/D5 system. Although the intersection is non-supersymmetric, we nevertheless expect that this provides a new way of counting the lower-dimensional black hole entropy.

Our construction is the simplest example beyond supergravities that admit analytical charged AdS black holes, with the RN-AdS black hole in general dimensions as a special solution. It suggests that there exist large classes of such theories with more dilatons and vector fields.  The explicit examples of the electrically-charged AdS black holes and the magnetic AdS$_{D-2}\times {\cal Y}^2$ vacua we have obtained provide a wealth of gravitational backgrounds to study the phenomenological aspects of the AdS/CFT correspondence.

\section*{Acknowledgement}

We are grateful to Eoin Colgain, Sera Cremonini, Mirjam Cveti\v c, Hai-Shan Liu, Chris Pope, Chiara Toldo and Justin V\'azquez-Poritz for useful discussions, and to ICTS-USTC for hospitality during the later stage of this work.  We are grateful to Justin V\'azquez-Poritz for pointing out an error in the super potential in section 5 in the earlier version. The research is supported in part by the NSFC grants 11175269 and 11235003.


\begin{thebibliography}{99}

\bibitem{Behrndt:1998jd}
  K.~Behrndt, M.~Cveti\v c and W.A.~Sabra,
{\it Nonextreme black holes of five-dimensional ${\cal N}=2$ AdS supergravity,}
  Nucl.\ Phys.\ B {\bf 553}, 317 (1999)
  [hep-th/9810227].

\bibitem{Duff:1999gh}
  M.J.~Duff and J.T.~Liu,
{\it Anti-de Sitter black holes in gauged ${\cal N} = 8$ supergravity,}
  Nucl.\ Phys.\ B {\bf 554}, 237 (1999)
  [hep-th/9901149].

\bibitem{tenauthor}
  M.~Cveti\v c, M.J.~Duff, P.~Hoxha, J.T.~Liu, H.~L\"u, J.X.~Lu, R.~Martinez-Acosta and C.N.~Pope {\it et al.},
{\it Embedding AdS black holes in ten-dimensions and eleven-dimensions,}
  Nucl.\ Phys.\ B {\bf 558}, 96 (1999)
  [hep-th/9903214].

\bibitem{Cvetic:1999un}
  M.~Cveti\v c, H.~L\"u and C.N.~Pope,
{\it Gauged six-dimensional supergravity from massive type IIA,}
  Phys.\ Rev.\ Lett.\  {\bf 83}, 5226 (1999)
  [hep-th/9906221].

\bibitem{Klemm:2012yg}
  D.~Klemm and O.~Vaughan,
{\it Nonextremal black holes in gauged supergravity and the real formulation of special geometry,}
  JHEP {\bf 1301}, 053 (2013)
  [arXiv:1207.2679 [hep-th]].

\bibitem{Klemm:2012vm}
  D.~Klemm and O.~Vaughan,
{\it Nonextremal black holes in gauged supergravity and the real formulation of special geometry II,}
  Class.\ Quant.\ Grav.\  {\bf 30}, 065003 (2013)
  [arXiv:1211.1618 [hep-th]].

\bibitem{Chong:2005hr}
  Z.-W.~Chong, M.~Cveti\v c, H.~L\"u and C.N.~Pope,
{\it General non-extremal rotating black holes in minimal five-dimensional gauged supergravity,}
  Phys.\ Rev.\ Lett.\  {\bf 95}, 161301 (2005)
  [hep-th/0506029].

\bibitem{Wu:2011gq}
  S.-Q.~Wu,
{\it General non-extremal rotating charged AdS black holes in five-dimensional $U(1)^3$ gauged supergravity: a simple construction method,}
  Phys.\ Lett.\ B {\bf 707}, 286 (2012)
  [arXiv:1108.4159 [hep-th]].

\bibitem{Chong:2004na}
  Z.-W.~Chong, M.~Cveti\v c, H.~L\" u and C.N.~Pope,
{\it Charged rotating black holes in four-dimensional gauged and ungauged supergravities,}
  Nucl.\ Phys.\ B {\bf 717}, 246 (2005)
  [hep-th/0411045].

\bibitem{Chow:2007ts}
  D.D.K.~Chow,
{\it Equal charge black holes and seven dimensional gauged supergravity,}
  Class.\ Quant.\ Grav.\  {\bf 25}, 175010 (2008)
  [arXiv:0711.1975 [hep-th]].

\bibitem{Chow:2008ip}
  D.D.K.~Chow,
{\it Charged rotating black holes in six-dimensional gauged supergravity,}
  Class.\ Quant.\ Grav.\  {\bf 27}, 065004 (2010)
  [arXiv:0808.2728 [hep-th]].

\bibitem{Rahmfeld:1995fm}
  J.~Rahmfeld,
{\it Extremal black holes as bound states,}
  Phys.\ Lett.\ B {\bf 372}, 198 (1996)
  [hep-th/9512089].

\bibitem{Tseytlin:1996bh}
  A.A.~Tseytlin,
{\it Harmonic superpositions of M-branes,}
  Nucl.\ Phys.\ B {\bf 475}, 149 (1996)
  [hep-th/9604035].

\bibitem{Lu:2011nga}
  H.~L\"u and Z.-L.~Wang,
{\it Pseudo-Killing spinors, pseudo-supersymmetric p-branes, bubbling and less-bubbling AdS Spaces,}
  JHEP {\bf 1106}, 113 (2011)
  [arXiv: 1103.0563 [hep-th]].

\bibitem{Lu:2011zx}
  H.~L\"u, C.N.~Pope and Z.-L.~Wang,
{\it Pseudo-supersymmetry, consistent sphere reduction and Killing spinors for the bosonic string,}
  Phys.\ Lett.\ B {\bf 702}, 442 (2011)
  [arXiv:1105.6114 [hep-th]].

\bibitem{Lu:2011vk}
  H.~L\"u and Z.-L.~Wang,
{\it Killing spinors for the bosonic String,}
  Europhys.\ Lett.\  {\bf 97}, 50010 (2012)
  [arXiv:1106.1664 [hep-th]].

\bibitem{Liu:2012jra}
  H.~Liu, H.~L\"u and Z.-L.~Wang,
{\it Killing spinors for the bosonic string and the Kaluza-Klein theory with scalar potentials,}
  Eur.\ Phys.\ J.\ C {\bf 72}, 1853 (2012)
  [arXiv:1106.4566 [hep-th]].

\bibitem{Lu:2011ku}
  H.~L\"u, C.N.~Pope and Z.-L.~Wang,
{\it Pseudo-supergravity extension of the bosonic string,}
  Nucl.\ Phys.\ B {\bf 854}, 293 (2012)
  [arXiv:1106.5794 [hep-th]].

\bibitem{Liu:2011ve}
  H.-S.~Liu, H.~L\"u, Z.-L.~Wang,
{\it Gauged Kaluza-Klein AdS pseudo-supergravity,}
  Phys.\ Lett.\ B {\bf 703}, 524 (2011)
  [arXiv:1107.2659 [hep-th]].

\bibitem{Chow:2011fh}
  D.D.K.~Chow,
{\it Single-rotation two-charge black holes in gauged supergravity,}
  arXiv:1108.5139 [hep-th].

\bibitem{Wu:2011zzh}
  S.-Q.~Wu,
{\it General rotating charged Kaluza-Klein AdS black holes in higher dimensions,}
  Phys.\ Rev.\ D {\bf 83}, 121502 (2011)
  [arXiv:1108.4157 [hep-th]].


\bibitem{Chan:1995fr}
  K.C.K.~Chan, J.H.~Horne and R.B.~Mann,
{\it Charged dilaton black holes with unusual asymptotics,}
  Nucl.\ Phys.\ B {\bf 447}, 441 (1995)
  [gr-qc/9502042].

\bibitem{stainless}
  H.~L\"u, C.N.~Pope, E.~Sezgin and K.S.~Stelle,
{\it Stainless super p-branes,}
  Nucl.\ Phys.\ B {\bf 456}, 669 (1995)
  [hep-th/9508042].

\bibitem{Ortin:2004ms}
  T.~Ortin,
{\it Gravity and strings,}
  Cambridge Unversity, Cambridge University Press, 2004

\bibitem{Almuhairi:2011ws}
  A.~Almuhairi and J.~Polchinski,
{\it Magnetic AdS$\times R^2$: supersymmetry and stability,}
  arXiv:1108.1213 [hep-th].

\bibitem{toda}
  H.~L\"u, C.N.~Pope and K.W.~Xu,
{\it Liouville and Toda solutions of M theory,}
  Mod.\ Phys.\ Lett.\ A {\bf 11}, 1785 (1996)
  [hep-th/9604058].

\bibitem{Lu:2013uia}
  H.~L\"u and W.~Yang,
{\it $SL(n,R)$-Toda black holes,}
  arXiv:1307.2305 [hep-th].

\bibitem{Arnowitt:1959ah}
  R.L.~Arnowitt, S.~Deser and C.W.~Misner,
{\it Dynamical structure and definition of energy in general relativity,}
  Phys.\ Rev.\  {\bf 116}, 1322 (1959).

\bibitem{CYII} M. Cveti\v c and D. Youm,
{\it Entropy of non-extreme charged rotating black holes in string theory},
Phys. Rev. {\bf D54} (1996) 2612, arXiv:hep-th/9603147.

  \bibitem{L}
  F. Larsen,
{\it A string model of black hole microstates},
Phys. Rev. {\bf D56} (1997) 1005, hep-th/9702153.

\bibitem{CLI}
M. Cveti\v c and F. Larsen,
{\it General rotating black holes in string theory: Greybody
factors and  event horizons},
Phys. Rev. {\bf D56}, 4994 (1997), hep-th/9705192.

\bibitem{CLII}
   M. Cveti\v c and F. Larsen,
{\it Greybody factors for rotating black holes in four dimensions},
Nucl. Phys. {\bf B506}, 107 (1997), hep-th/9706071.

\bibitem{Chen:2011in}
  Y.-X.~Chen, H.~L\"u and K.~-N.~Shao,
{\it Linearized modes in extended and critical gravities,}
  Class.\ Quant.\ Grav.\  {\bf 29}, 085017 (2012)
  [arXiv:1108.5184 [hep-th]].

\bibitem{Lu:2011qx}
  H.~L\"u and K.~-N.~Shao,
{\it Solutions of free higher spins in AdS,}
  Phys.\ Lett.\ B {\bf 706}, 106 (2011)
  [arXiv:1110.1138 [hep-th]].

\bibitem{Townsend:1983kk}
  P.K.~Townsend and P.~van Nieuwenhuizen,
{\it Gauged seven-dimensional supergravity,}
  Phys.\ Lett.\ B {\bf 125}, 41 (1983).

\bibitem{Romans:1985tw}
  L.J.~Romans,
{\it The $F(4)$ gauged supergravity in six-dimensions,}
  Nucl.\ Phys.\ B {\bf 269}, 691 (1986).

\bibitem{lpsdilatonic}
  H.~L\"u, C.N.~Pope, E.~Sezgin and K.S.~Stelle,
{\it Dilatonic $p$-brane solitons,}
  Phys.\ Lett.\ B {\bf 371}, 46 (1996)
  [hep-th/9511203].

\bibitem{Bremer:1998zp}
  M.S.~Bremer, M.J.~Duff, H.~L\"u, C.N.~Pope and K.S.~Stelle,
{\it Instanton cosmology and domain walls from M theory and string theory,}
  Nucl.\ Phys.\ B {\bf 543}, 321 (1999)
  [hep-th/9807051].

\bibitem{Liu:2012ed}
  H.~Liu, H.~L\"u and Z.-L.~Wang,
{\it $f(R)$ theories of supergravities and pseudo supergravities,}
  JHEP {\bf 1204}, 072 (2012)
  [arXiv:1201.2417 [hep-th]].

\bibitem{Chen:2005zj}
  W.~Chen, H.~L\"u and C.~N.~Pope,
  {\it Mass of rotating black holes in gauged supergravities,}
  Phys.\ Rev.\ D {\bf 73}, 104036 (2006)
  [hep-th/0510081].

\bibitem{Ashtekar:1984zz}
  A.~Ashtekar and A.~Magnon,
{\it Asymptotically anti-de Sitter space-times,}
  Class.\ Quant.\ Grav.\  {\bf 1}, L39 (1984).

\bibitem{Ashtekar:1999jx}
  A.~Ashtekar and S.~Das,
{\it Asymptotically Anti-de Sitter space-times: conserved quantities,}
  Class.\ Quant.\ Grav.\  {\bf 17}, L17 (2000)
  [hep-th/9911230].

\bibitem{Romans:1991nq}
  L.J.~Romans,
{\it Supersymmetric, cold and lukewarm black holes in cosmological Einstein-Maxwell theory,}
  Nucl.\ Phys.\ B {\bf 383}, 395 (1992)
  [hep-th/9203018].

\bibitem{AlonsoAlberca:2000cs}
  N.~Alonso-Alberca, P.~Meessen and T.~Ortin,
{\it Supersymmetry of topological Kerr-Newman-Taub-NUT-AdS space-times,}
  Class.\ Quant.\ Grav.\  {\bf 17}, 2783 (2000)
  [hep-th/0003071].

\bibitem{Lin:2004nb}
  H.~Lin, O.~Lunin and J.M.~Maldacena,
{\it Bubbling AdS space and 1/2 BPS geometries,}
  JHEP {\bf 0410}, 025 (2004)
  [hep-th/0409174].

\bibitem{Chong:2004ce}
  Z.-W.~Chong, H.~L\"u and C.~N.~Pope,
{\it BPS geometries and AdS bubbles,}
  Phys.\ Lett.\ B {\bf 614}, 96 (2005)
  [hep-th/0412221].

\bibitem{Gutowski:2004ez}
  J.B.~Gutowski and H.S.~Reall,
{\it Supersymmetric AdS$_5$ black holes,}
  JHEP {\bf 0402}, 006 (2004)
  [hep-th/0401042].

\bibitem{Cacciatori:2009iz}
  S.L.~Cacciatori and D.~Klemm,
{\it Supersymmetric AdS$_4$ black holes and attractors,}
  JHEP {\bf 1001}, 085 (2010)
  [arXiv:0911.4926 [hep-th]].

\bibitem{Toldo:2012ec}
  C.~Toldo and S.~Vandoren,
{\it Static nonextremal AdS$_4$ black hole solutions,}
  JHEP {\bf 1209}, 048 (2012)
  [arXiv:1207.3014 [hep-th]].

\bibitem{cvgipo} M.~Cveti\v c, G.W.~Gibbons and C.N.~Pope,
{\it Universal area product formulae for rotating and charged black holes
in four and higher dimensions},
Phys.\ Rev.\ Lett.\  {\bf 106}, 121301 (2011),
arXiv:1011.0008 [hep-th].

\bibitem{Gubser:2012yb}
  S.S.~Gubser and J.~Ren,
{\it Analytic fermionic Green's functions from holography,}
  Phys.\ Rev.\ D {\bf 86}, 046004 (2012)
  [arXiv:1204.6315 [hep-th]].

\bibitem{Bhattacharya:2012zu}
  J.~Bhattacharya, S.~Cremonini and A.~Sinkovics,
{\it On the IR completion of geometries with hyperscaling violation,}
  JHEP {\bf 1302}, 147 (2013)
  [arXiv:1208.1752 [hep-th]].

\bibitem{Das:1977pu}
  A.~Das, M.~Fischler and M.~Rocek,
  Phys.\ Rev.\ D {\bf 16}, 3427 (1977).

\bibitem{Gibbons:1985ac}
  G.W.~Gibbons and D.L.~Wiltshire,
{\it Black holes in Kaluza-Klein theory,}
  Annals Phys.\  {\bf 167}, 201 (1986)
  [Erratum-ibid.\  {\bf 176}, 393 (1987)].

\bibitem{Cvetic:1999au}
  M.~Cveti\v c, H.~L\"u and C.N.~Pope,
{\it Four-dimensional ${\cal N}=4, SO(4)$ gauged supergravity from $D = 11$,}
  Nucl.\ Phys.\ B {\bf 574}, 761 (2000)
  [hep-th/9910252].

\bibitem{Cucu:2003bm}
  S.~Cucu, H.~L\"u and J.F.~V\'azquez-Poritz,
{\it A Supersymmetric and smooth compactification of M theory to AdS$_5$,}
  Phys.\ Lett.\ B {\bf 568}, 261 (2003)
  [hep-th/0303211].

\bibitem{Cucu:2003yk}
  S.~Cucu, H.~L\"u and J.F.~V\'azquez-Poritz,
{\it Interpolating from AdS$_{D-2} \times S^2$ to AdS$_{D}$,}
  Nucl.\ Phys.\ B {\bf 677}, 181 (2004)
  [hep-th/0304022].

\bibitem{Almuhairi:2010rb}
  A.~Almuhairi,
{\it AdS$_3$ and AdS$_2$ magnetic brane solutions,}
  arXiv:1011.1266 [hep-th].

\bibitem{Anderson:2011cz}
  M.T.~Anderson, C.~Beem, N.~Bobev and L.~Rastelli,
{\it Holographic uniformization,}
  Commun.\ Math.\ Phys.\  {\bf 318}, 429 (2013)
  [arXiv:1109.3724 [hep-th]].

\bibitem{Maldacena:2000mw}
  J.M.~Maldacena and C.~Nunez,
{\it Supergravity description of field theories on curved manifolds and a no go theorem,}
  Int.\ J.\ Mod.\ Phys.\ A {\bf 16}, 822 (2001)
  [hep-th/0007018].

\bibitem{Cacciatori:1999rp}
  S.~Cacciatori, D.~Klemm and D.~Zanon,
{\it $w_\infty$ algebras, conformal mechanics, and black holes,}
  Class.\ Quant.\ Grav.\  {\bf 17}, 1731 (2000)
  [hep-th/9910065].

\bibitem{Klemm:2000nj}
  D.~Klemm and W.A.~Sabra,
{\it Supersymmetry of black strings in $D = 5$ gauged supergravities,}
  Phys.\ Rev.\ D {\bf 62}, 024003 (2000)
  [hep-th/0001131].


\end{thebibliography}
\end{document}